\documentclass{elsart}

\usepackage{subfigure,graphicx,epsfig}
\usepackage[boxed]{algorithm}
\usepackage{algorithmic}
\usepackage{rotating}

\usepackage{amssymb}

\newcommand{\iotabar}{\raisebox{-1pt}{$\mathchar'40$}\mkern-5.43mu\iota}

\begin{document}

\begin{frontmatter}



\title{A data mining algorithm for automated characterisation of fluctuations in multichannel timeseries}


\author{D. G. Pretty\corauthref{cor1}},
\ead{david.pretty@anu.edu.au}
\author{B. D. Blackwell}
\ead{boyd.blackwell@anu.edu.au}
\corauth[cor1]{Corresponding author}

\address{Plasma Research Laboratory, Research School of Physical Sciences and Engineering, Australian National University, Canberra ACT 0200, Australia}

\begin{abstract}
We present a data mining technique for the analysis of multichannel oscillatory timeseries data and show an application using poloidal arrays of magnetic sensors installed in the H-1 heliac. The procedure is highly automated, and scales well to large datasets. The timeseries data is split into short time segments to provide time resolution, and each segment is represented by a singular value decomposition (SVD). By comparing power spectra of the temporal singular vectors, singular values are grouped into subsets which define \emph{fluctuation structures}. Thresholds for the normalised energy of the fluctuation structure and the normalised entropy of the SVD are used to filter the dataset. We assume that distinct classes of fluctuations are localised in the space of phase differences $\Delta\psi(n,n+1)$ between each pair of nearest neighbour channels. An expectation maximisation clustering algorithm is used to locate the distinct classes of fluctuations, and a \emph{cluster tree} mapping is used to visualise the results.

\end{abstract}

\end{frontmatter}

\section{Introduction}
\label{sec:introduction}

The motivation for the present work arose from the analysis of fluctuations in magnetically confined plasma during parameter scans in the H-1 flexible heliac \cite{Hamberger:1990,Harris:2004}. The H-1 heliac is a three field-period helical axis stellarator \cite{Blackwell:2001} with major radius $R = 1\,$m, minor radius $\langle r \rangle = 0.2\,$m and a finely tunable magnetic geometry. 

Experimental scans through plasma configurations via the geometric parameter $\kappa_h$, which controls the rotational  transform $\iotabar$ (twist of the magnetic field lines) and shear  $\iotabar^\prime$(radial derivative of rotational transform), have produced diverse spectra of magnetohydrodynamic (MHD) activity. The MHD activity is recorded via two toroidally separated poloidal arrays of Mirnov coils (induction solenoids) which sample $dB/dt$ locally. In the example dataset presented here, 28 Mirnov coils are used for 92 distinct plasma configurations, resulting in more than 100,000 short time Fourier spectra.

The data mining process used to reduce this dataset is described in the following sections. In section \ref{sec:preprocessing} we explain the preprocessing stage, which includes filtering and mapping into a high dimensional phase space. In section \ref{sec:clustering} the clustering algorithm for distinguishing classes of fluctuations is described, followed by a demonstration of a visualisation procedure. A discussion of some important aspects of the procedure follows in section \ref{sec:discussion}.

\section{Preprocessing}
\label{sec:preprocessing}
\subsection{Data preparation}
We assume that each set of timeseries data can be represented as a $N_c\times N_s$ matrix:
\begin{equation}
\mathcal{S} = \left(\begin{array}{ccccc} s_0(t_0) & s_0(t_0+\tau) & s_0(t_0+2\tau) & \ldots & s_0(t_0+N_s\tau) \\ s_1(t_0) & s_1(t_0+\tau) & s_1(t_0+2\tau) & \ldots & s_1(t_0+N_s\tau) \\ \vdots & \vdots & \vdots & \ddots & \vdots \\ s_{N_c}(t_0) & s_{N_c}(t_0+\tau) & s_{N_c}(t_0+2\tau) & \ldots & s_{N_c}(t_0+N_s\tau) \end{array}\right)
\end{equation}
where $\tau$ is the inverse of the sampling frequency, $N_c$ is the number of channels and $N_s$ is the number of samples. In our example dataset, the signal amplitudes depend on the plasma-coil distance which is a function of the plasma shape (magnetic configuration) controlled by $\kappa_h$. To reduce any configurational bias on $\mathcal{S}$ we normalise each channel to its variance. 

To achieve time resolution $\Delta t$, we split $\mathcal{S}$ into short time segments $S$ with shape $N_c\times N_s^\prime$, where $N_s^\prime = \Delta t/\tau$. We also assume there are an arbitrary number of $\mathcal{S}$ relating to the same system, e.g.: an experiment repeated under different conditions. At this stage there is no need to distinguish between the $S$ from different $\mathcal{S}$, although we implicitly retain sufficient information to map the $S$ back to their original parameter sets.

\subsection{The singular value decomposition}
\begin{figure}
 	\centering
 	\includegraphics[scale=0.5]{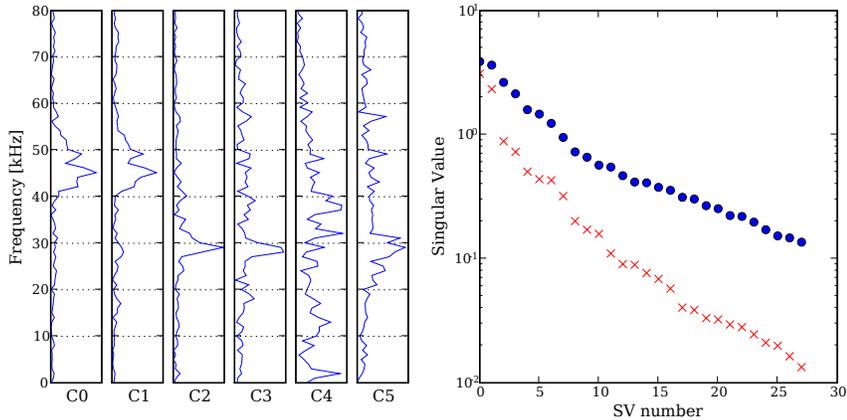}
 	\caption{Example of chronos power spectra and singular values. Singular values from both normalised (o) and unnormalised (x) $S$ are shown. $C0, C1, \ldots, C5$ denote the chronos from the normalised singular value $0,1,\ldots, 5$. There are two distinct modes, one at $f\sim 45\,$kHz described by SV0 and SV1; the other, weaker, signal is at $f\sim 29\,$kHz and is described by SV2 and SV3. The data is from H1 shot \#58122 at $31 < t < 32\,$ms.}
 	\label{fig:svdexample}
\end{figure}

Each $S$ is represented by a singular value decomposition (SVD) \cite{Dudok:1994}
\begin{equation}
S = UAV^*
\end{equation}
where the columns of $U$ and $V$ contain the spatial (topo) and temporal (chrono)  singular vectors respectively, $V^*$ denotes the conjugate transpose of $V$, and the diagonal elements of $A$ are the $N_a=\min(N_c,N_s^\prime)$ non-negative singular values. The set of topos (chronos) are an orthonormal basis of $\mathbf{R}^{N_c (N_s^\prime)}$. The convention is for the singular values to be sorted in decreasing monotonic order meaning that $A$ is independent of the ordering of the channels within $S$. Shown in figure \ref{fig:svdexample} are singular values from a typical H1 Mirnov dataset. From the chronos power spectra we see that there are two dominant modes, each with two singular values suggesting they are both travelling waves, as discussed below. We also see that the variance-normalisation of each channel degrades the signal to noise ratio of the system, which can also be described in terms of the normalised entropy.

We calculate the normalised entropy $H$ of the singular values $a_k$ in $A$:
\begin{equation}
H = \frac{-\sum_{k=1}^{N_a}p_k\log{p_k}}{\log{N_a}},
\end{equation}
where $p_k$ is the dimensionless energy:
\begin{equation}
p_k = \frac{a_k^2}{E},\quad E = \sum_{k=1}^{N_a}a_k^2.
\label{eqn:pE}
\end{equation}
The low entropy case ($H \rightarrow 0$) occurs when the system is well ordered. To some extent the scalar quantity $H$ can be used as a measure of how physically interesting the signals in $S$ are  without any further investigation into the structure of $S$, though care must be taken with this interpretation. A standing wave in a system with no noise has only one non-zero singular value $a_0=1$ giving $H=0$, whereas a travelling wave requires two singular values so $H>0$.

\subsection{Fluctuation structures}

Recognising that a travelling wave structure consists of a pair of singular values which naturally belong together we group similar singular values, defining a \emph{fluctuation structure} $\alpha$ as a subset of singular values which have chronos with similar power spectra. 
We measure the similarity between two chronos $c_1$ and $c_2$ with the normalised average of the cross-power spectrum  $\gamma_{c_1,c_2}$:
\begin{equation}
\gamma_{c_1,c_2} = \frac{G(c_1,c_2)^2}{G(c_1,c_1)G(c_2,c_2)},
\end{equation}
where $G(a,b) = \langle | \mathcal{F}(a)\mathcal{F}^*(b)| \rangle$, $\mathcal{F}$ is the Fourier transform, and $\langle\ldots\rangle$ represents the spectral average.

When allocating singular values to fluctuation structures, the observation:
\begin{equation}
\gamma_{a,b} > \gamma_{min}\quad \mathrm{and} \quad \gamma_{a,c} > \gamma_{min} \quad \nRightarrow \quad \gamma_{b,c} > \gamma_{min},
\end{equation}
suggests that we should not simply seek to require $\gamma_{a,b} > \gamma_{min}$ for each pair of singular values $a,b$ within a structure, instead we follow the process in algorithm 1. In so doing we therefore require that each constituent singular vector has sufficient $\gamma$ with the dominant singular vector of the structure. 
\begin{algorithm}
\begin{algorithmic}
\WHILE{Number of unallocated singular values $>0$}
\STATE Define a new fluctuation structure as an empty set of singular values: $\alpha_i = \{\}$
\STATE Denote the largest unallocated singular value by $a_\xi$
\FOR{Every unallocated singular value $a_\zeta$}
\IF{$\gamma_{\zeta,\xi} > \gamma_{min}$}
\STATE Allocate  $a_\zeta$ to fluctuation structure $\alpha_i$
\ENDIF
\ENDFOR
\ENDWHILE 
\end{algorithmic}
\caption{Building fluctuation structures $\alpha_i$ from singular values $a_j$. The largest unallocated singular value $a_\xi$ will always be allocated to $\alpha_i$ because $\gamma_{\xi,\xi} = 1$. }
\end{algorithm}

\begin{figure}
 	\centering
 	\includegraphics[scale=0.5]{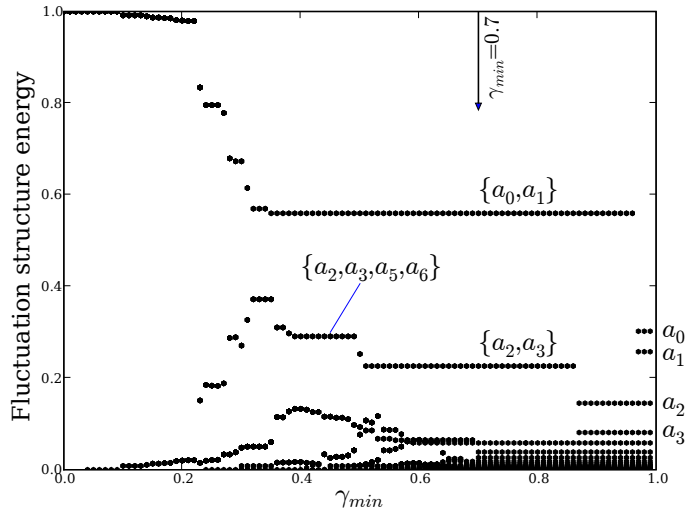}
 	\caption{The possible fluctuation structure groupings according to their energies as defined by algorithm 1 through the range of $\gamma_{min}$. The dataset is the same as in figure \ref{fig:svdexample}. We see that $\gamma_{min} \lesssim 0.4$ allows unrelated singular values to be included within a fluctuation structure, whereas with $\gamma_{min} > 0.87$ algorithm 1 will not recognise the similarity between $a_2$ and $a_3$.}
 	\label{fig:gammaenergy}
\end{figure}

Various possible fluctuation structures for the dataset of figure \ref{fig:svdexample} are shown in figure \ref{fig:gammaenergy} as a function of the threshold value $\gamma_{min}$. At $\gamma_{min} = 0$, all singular values are grouped together as a single fluctuation structure, while at $\gamma_{min} = 1$ each fluctuation structure contains one singular value. The key features are the two fluctuation structures $\alpha_0 = \{a_0,a_1\}$ and $\alpha_1 = \{a_2,a_3\}$ which coexist for $0.50 < \gamma_{min} < 0.87$. After application of such analysis to a suitably sized sample of short time segments, a threshold of $\gamma_{min} = 0.7$ was found to be appropriate for our dataset.

\subsection{Data filtering}

Filters are applied to the dataset in order to reduce its size and to remove noise. Two values which can be used to quantify the quality of the data are the normalised energy $p$ and normalised entropy $H$. The normalised energy $p$ of a fluctuation structure is defined as the sum of the normalised energies of its constituent singular values from equation \ref{eqn:pE}. The nature of these thresholds is quite different; $H$ thresholds will act upon the entire short time segments, whereas $p$ thresholds affect individual fluctuation structures.

\begin{figure}
 	\centering
 	\includegraphics[scale=0.25]{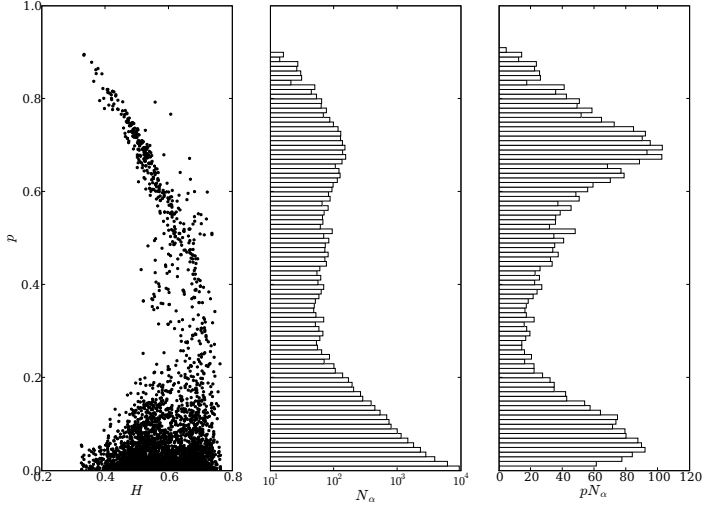}
 	\caption{A $10\%$ random sample of a dataset. The left panel shows $p$ and $H$ for the fluctuation structures. The middle panel shows the number of fluctuation structures $N_\alpha$ within $\delta p = 0.01$. The right panel shows $pN_\alpha$, which is effectively the density of normalised energy; while this is not physically meaningful because the normalisation factor is dependent on short time segment, it is a useful guide to the energy distribution among fluctuation structures.}
 	\label{fig:HP_with_p_int}
\end{figure}

Using a normalised energy threshold value $p^\prime$ allows filtering out of low energy noise. As seen in our example dataset (figure \ref{fig:HP_with_p_int}), there appears to be a clear distinction between higher energy fluctuations ($p\gtrsim 0.6$) and lower energy noise ($p\lesssim 0.2$). The use of $H$ thresholds is not always appropriate, especially if spectra are present with several distinct fluctuations; in such cases if the dataset needs to be reduced in size it is preferable to simply use a random subset of the data. Entropy filtering is generally more useful when a significant number of $S$ contain only noise.

An alternative to using a hard $p$ threshold uses an energy threshold defined as a fraction of the possible range of normalised energy for a given singular value, before constructing fluctuation structures. This method is more sensitive to modes which have reduced $p_k$ due to the coexistence of other modes. The $n^\mathrm{th}$ largest singular value in a given short time segment has, by definition, a maximal normalised energy of $1/n$. We then apply a factor $p^\ast$, where $0<p^\ast<1$, and, starting with the smallest singular value, retain singular values with $p_n\geq p^\ast/n$. Any point retained brings in all larger singular values from the same time segment, trumping the $p_n\geq p^\ast/n$ condition for lower $n$. The requirement for bringing in larger singular values is easily understood by considering the case of a mode having two singular values of almost equal energy, it is possible for the lower energy value to exceed the threshold with the higher value below the threshold.

\subsection{Mapping of fluctuation structures into $\Delta\psi$-space}
\label{sec:mappingfs}

We regard each fluctuation structure as a point in the space $[-\pi,\pi]^{N_c}$,  an $N_c$--dimensional torus of length $2\pi$ which we will call $\Delta \psi$-space. In this application $\psi$ represents the electrical phase of the reconstructed fluctuation structure at the positions of the coils. Fluctuation structures which are close in $\Delta \psi$-space can be considered the same type. This interpretation arises from the expectation that the possible waves have various phase velocities and mode numbers due to the periodic boundary conditions of the physical plasma torus in which they propagate. It is also applicable to the more general case where waves are spatially localised within the system and do not have well defined mode numbers.

For each fluctuation structure $\alpha_l$ we take the inverse SVD to get $S_l$:
\begin{equation}
S_l = UA_lV^*, 
\end{equation}
where the elements of $A$ not in $\alpha_l$ are set to zero to form $A_l$. The rows of the matrix $S_l$ contain the timeseries relating to  $\alpha_l$ for each channel. In general, the power spectra of the topos in $\alpha_l$ are peaked around a single frequency $\omega_l$. The phase differences $\Delta \psi_{a,b}(\omega=\omega_l)$ between channels $a$ and $b$ evaluated at $\omega=\omega_l$ are used to define the coordinates in $\Delta \psi$-space. Using phase differences between each pair of channels would result in a $\frac{1}{2}N_c(N_c-1)$-dimensional space; instead we use the $N_c-$dimensional space of only nearest neighbour channels. Note that in our example dataset the actual phase difference between channels depends on $\kappa_h$ so we map the phase differences to a coordinate system which is independent of $\kappa_h$, namely the $\kappa_h$-averaged magnetic angles of the Mirnov coils.

\subsection{An overview of the preprocessed dataset}

\begin{figure}
 	\centering
 	\includegraphics[scale=0.5]{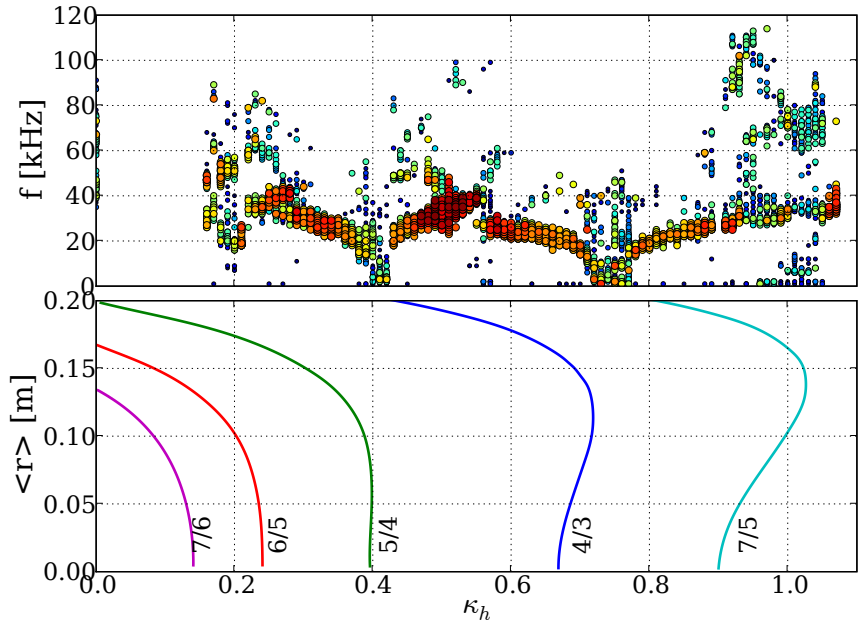}
 	\caption{The preprocessed dataset: In the upper panel, fluctuation structures are mapped to frequency and the magnetic geometry parameter $\kappa_h$, marker size and colour are proportion to the normalised energy of the fluctuation structure. In the lower panel, the average minor radial $\langle r \rangle$ locations of low order rational magnetic surfaces are shown.}
 	\label{fig:kappafreq}
\end{figure}

As an overview of the preprocessed dataset, figure \ref{fig:kappafreq} shows the fluctuation structures with energy $p>0.2$ mapped to the magnetic geometry parameter $\kappa_h$; the radial location of low-order rational magnetic surfaces, $\iotabar = n/m$, are also shown. These rational surfaces are important as fluctuations with toroidal and poloidal mode numbers $n$ and $m$ respectively can resonate with the twisted field lines. The main features of the fluctuation spectra are the resonances about $\kappa_h = 0.4$ and $\kappa_h = 0.76$, related to the $\iotabar = 5/4$ and $\iotabar=4/3$ surfaces respectively. We expect that any automated process used to locate distinct types of fluctuations would identify these features, and hopefully find some less obvious features. Indeed in section \ref{sec:discussion} it can be seen that these
two features are the first to be distinguished by the following clustering algorithm.

\section{Clustering}
\label{sec:clustering}

We aim to discover any underlying lower-dimensional model of the dataset; that is, groups of fluctuation structures which are similar throughout some range of short time segments. 
As discussed in section \ref{sec:mappingfs}, we assume that a class of fluctuations is localised in the $N_c$-dimensional $\Delta\psi$-space. For example, it is simple to understand such localisation in terms of a simple cylindrical geometry with equidistant poloidal measurements, where each mode with poloidal mode number $m$ will be located at $\Delta\psi = 2\pi m/N_c$ in each dimension. However, we assume a generalised case in which the fluctuation may have arbitrary, including localised, structure.

Many different types of clustering algorithms exist; here we use the expectation maximisation (EM) algorithm which is a method for estimating the most likely  values of latent variables in a probabilistic model\cite{Dempster:1977}. Here we assume that each type of fluctuation can be described by a $N_c$-dimensional Gaussian distribution in $\Delta\psi$  space. The latent variables are the mean $\mu_i$ and standard deviation $\sigma_i$ for each cluster $i$, where $i = 1,2,3,\ldots,N_{Cl}$ and $N_{Cl}$ is the number of clusters. Given the initial conditions, in the form of random initial $\mu_i$ and $\sigma_i$ values for a prescribed number of clusters, the EM algorithm consists of two steps which repeat until a convergence criterion is met. Firstly, the expectation step assigns to each datapoint a probability, or expectation value, of belonging to each cluster which is calculated with the Gaussian distribution function. Secondly, $\mu_i$ and $\sigma_i$ are recalculated using the new expectation values as weight factors.

The 10-fold cross-validated log-likelihood ratio is used as a measure of how well the cluster assignments fit the data. The cross-validation process involves partitioning the dataset into random subsamples and comparing results from each subset to avoid oversensitivity to outliers in the data. The likelihood is the conditional probability of obtaining the cluster means and standard deviations given the observed data. Because the EM algorithm can only guarantee a local maximum in likelihood we use a Monte Carlo approach, with multiple repetitions with different randomised initial conditions for each $N_{Cl}$.

\subsection{Visualisation}
\label{sec:vis}

The identification of the correct number of clusters, or of those which are important, is a task that is by no means trivial to automate. We have found inspection of a dendrogram, or \emph{cluster tree},  mapping to be a practical method for identifying the important clusters. The cluster tree displays clusters for each $N_{Cl}$ below some maximum value $N_{Cl,max}$, with all clusters for a given $N_{Cl}$ forming a single tree level. Each child cluster is mapped to the cluster on the parent level with which it has the largest fraction of common datapoints. Cluster branches which do not fork over a significant range of $N_{Cl}$ are deemed to be well defined, and the point where well defined clusters start to break up suggests that $N_{Cl}$ is too high. While this procedure is clearly a subjective one, it is effective and does not depend on the type of clustering algorithm used.

\begin{sidewaysfigure}
 	\centering
 	\includegraphics[height=0.7\textwidth]{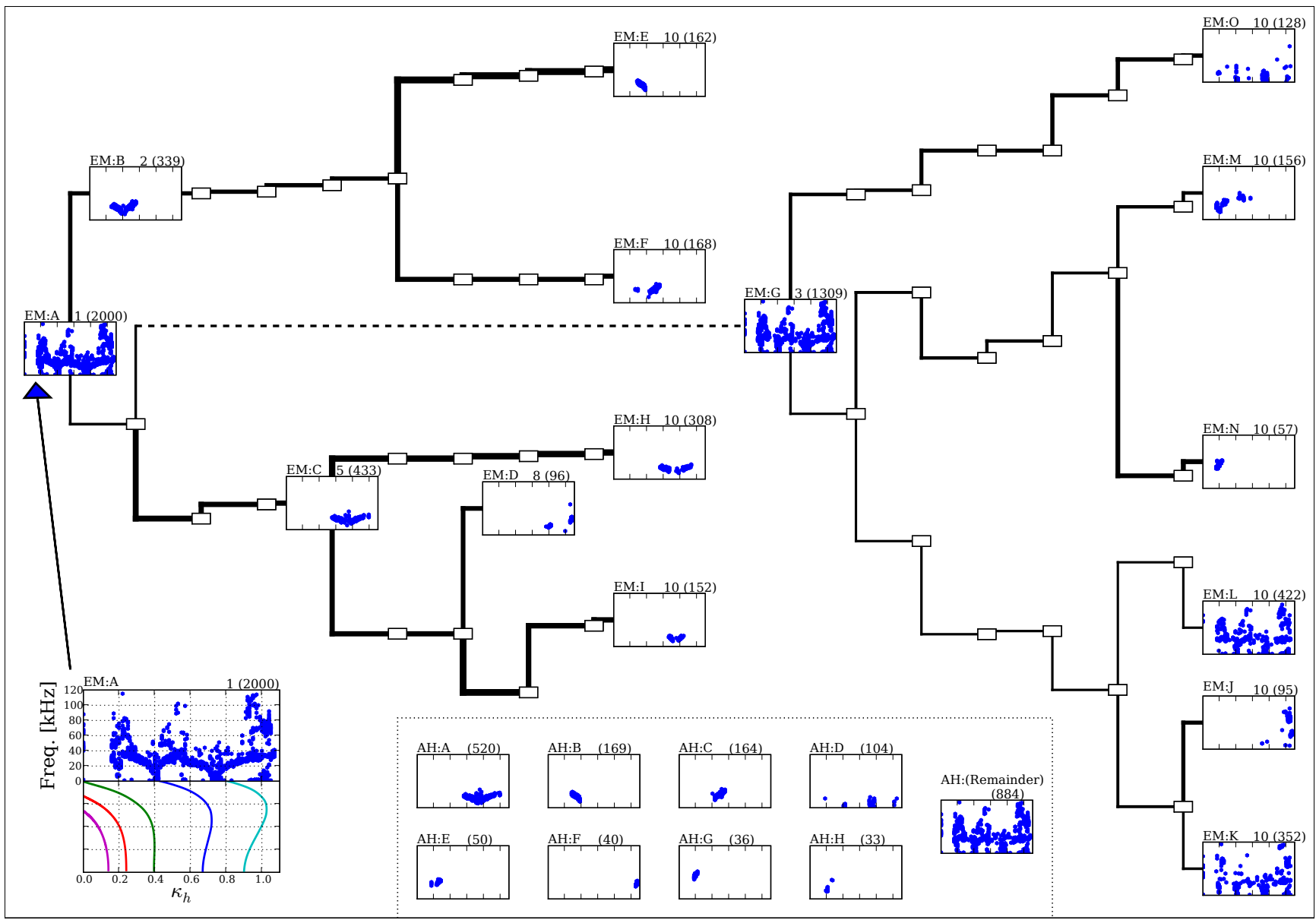}
 	\caption{Cluster tree of the example dataset. The figure in the bottom left corner is equivalent to figure \ref{fig:kappafreq}; its upper panel shows the fluctuation structures mapped to $f$ and $\kappa_h$, the numbers 1 (2000) at the top right are the tree level, $N_{Cl}$ , and cluster population respectively, EM:A is a cluster label used for reference. For clarity, only a subset of clusters within the tree have their contents displayed and EM:G has been displaced to prevent overlap. Vertical parent-child distance is proportional to the distance between cluster means, while line thickness is inversely proportional to the Gaussian width of the cluster. Several clusters produced by the agglomerative hierarchical (AH: labels) method are also shown, these are essentially equivalent to the $N_{Cl} = 10$ level EM clusters, see table \ref{arr:EM_AH_matrix_tgt10ms} for comparison.}
 	\label{fig:eg_cluster_tree}
 \end{sidewaysfigure}

The cluster tree for our example dataset is shown in figures \ref{fig:eg_cluster_tree}. Each cluster has been defined only by its phase structure $\psi$ and mapped back to $\kappa_h$ and frequency $f = (2\pi)^{-1}\omega_l$. The base of the tree ($N_{Cl}=1$) shows all the data within a single cluster (EM:A); as we climb up the tree different classes of fluctuation are isolated. For example, the branch starting at cluster EM:B contains fluctuations with toroidal mode number $n = 5$ and poloidal mode number $m=4$ which occur at configurations near the $\iotabar=5/4$ resonance ($\kappa_h \simeq 0.4$). Similarly, the branch containing cluster EM:C is due to the $\iotabar=4/3$ resonance near $\kappa_h = 0.75$. Other clusters include fluctuations which occur at higher order resonances, as well as low frequency $n,m=0$ modes (cluster EM:O branch) and weakly defined residual clusters (EM:K and EM:L) which would be resolved at a higher level of the tree than is shown here. 

\begin{figure}
 \centering
 \mbox{\subfigure[Cluster 47]{\epsfig{figure=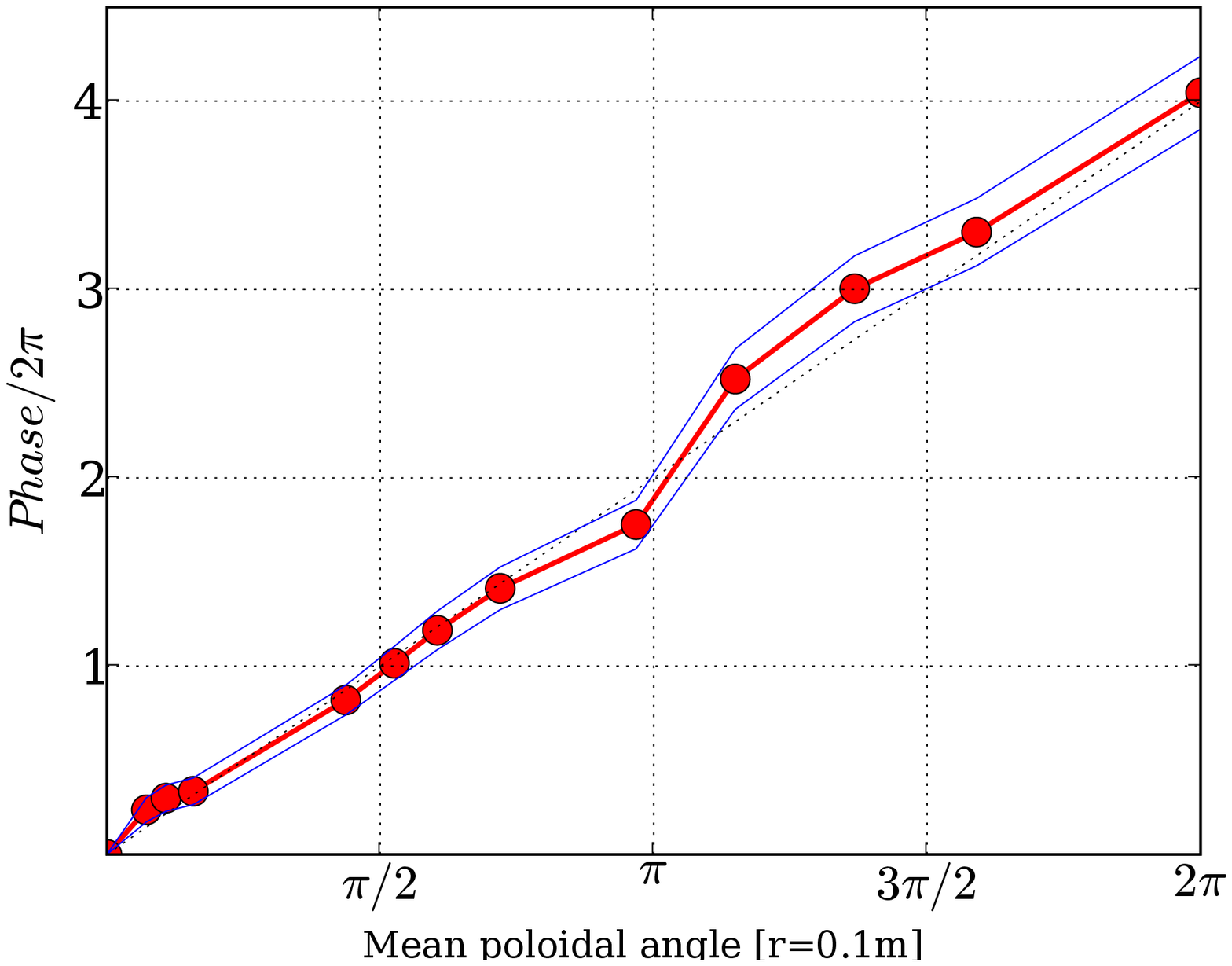,width=.37\textwidth}\label{fig:cl47phase}}\quad
 \subfigure[Cluster 48]{\epsfig{figure=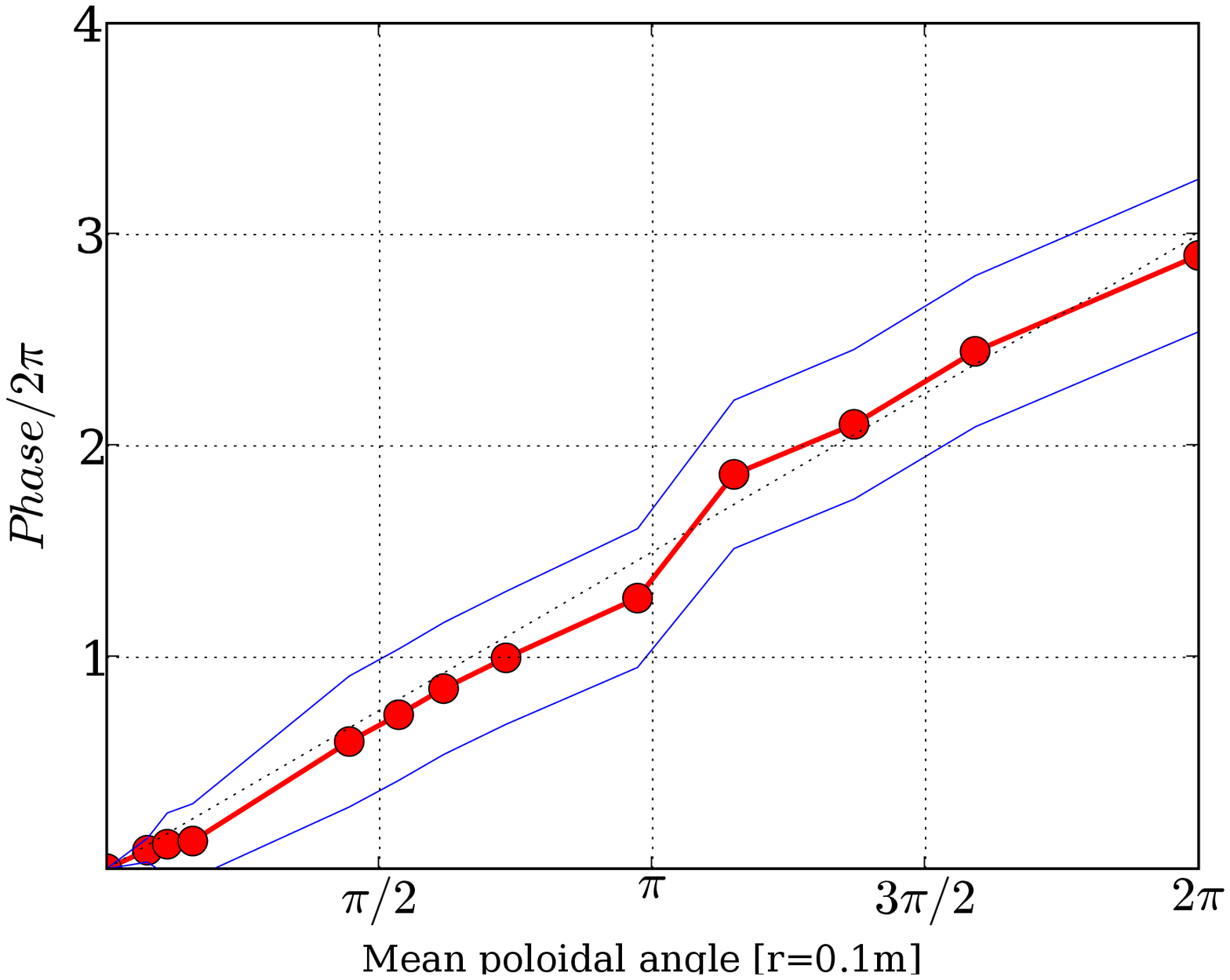,width=.37\textwidth}\label{fig:cl48phase}}\quad
 \subfigure[cluster 46]{\epsfig{figure=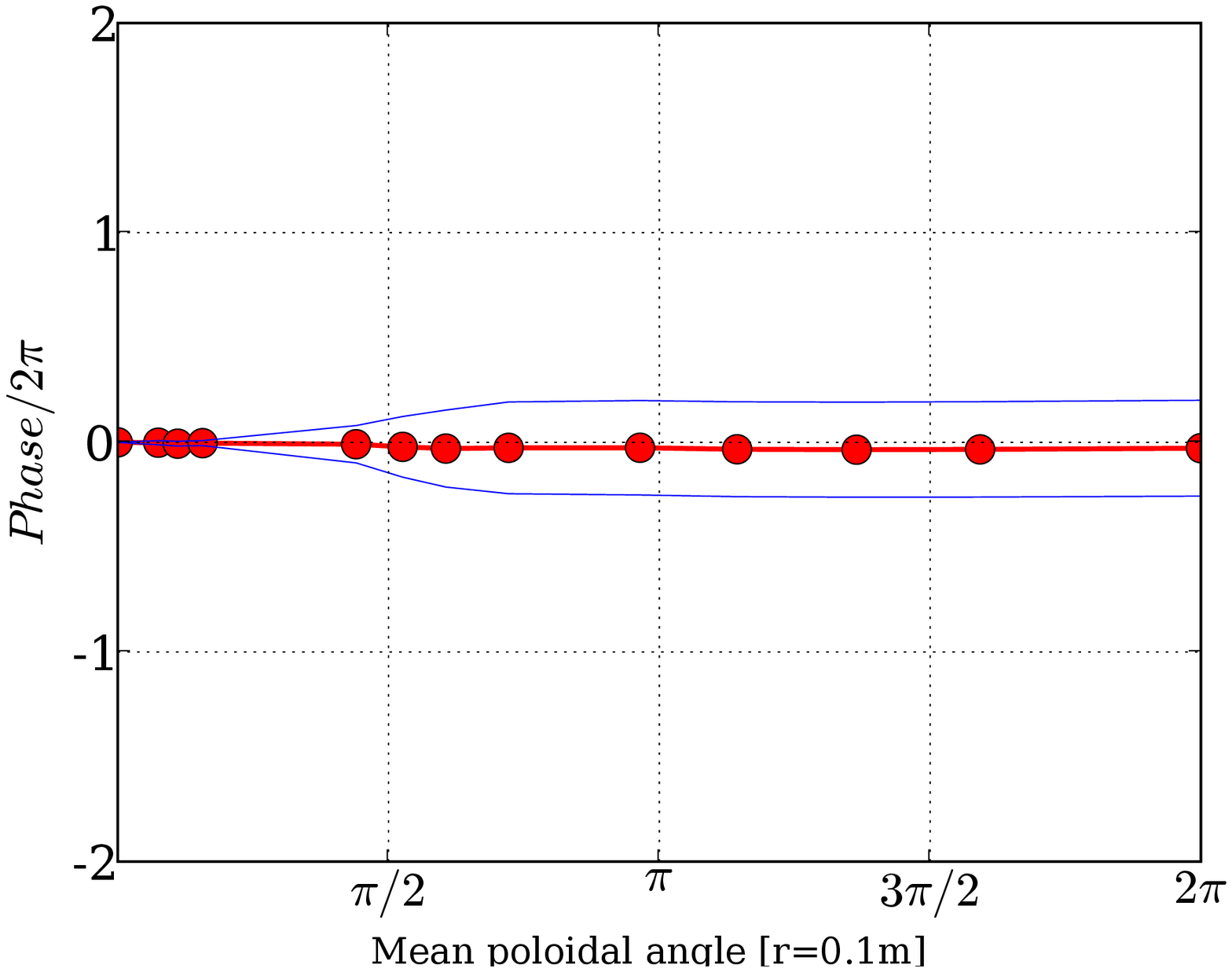,width=.37\textwidth}\label{fig:cl46phase}}}
 \caption{Variation of phase around one of the poloidal Mirnov arrays, plotted against mean magnetic poloidal coordinate. The centre line is the cumulative mean phase of the coil pairs, with standard deviation shown above and below. The mode numbers shown here are supported by Fourier analysis of the data.}
 \label{fig:poloidalphase}
 \end{figure}

Shown in figure \ref{fig:poloidalphase} are poloidal phase-angle plots for a single poloidal Mirnov array. The centre line corresponds to the cumulative mean phase of the coil pairs. The lines above and below are the cumulative cluster standard deviations of the coil pairs, where  $\sigma_{1,n}^2 = \sum_{j=1}^{n-1}\sigma_{j,j+1}^2$ for $\psi_{1,n} = \sum_{j=1}^{n-1}\Delta\psi_{j,j+1}$. Here, the magnetic angles have been evaluated for a flux surface at $r=0.1\,$m to better represent the broad radial structure expected for these modes.

\section{Discussion}
\label{sec:discussion}

The data mining algorithm presented here is potentially useful in numerous other domains where spatio-temporal data is used. However, there is a limitation to the nature of the fluctuations amenable to this analysis due to the SVD. The SVD is not effective in distinguishing different modes coexisting with the same frequency or spatial structure because the modes would share a chrono or topo, whereas the SVD requires orthogonal components to distinguish modes. The assumption that such coexisting modes are not present is also important in assigning a single frequency $\omega_l$ to a fluctuation structure, i.e.: two modes with the same spatial structure will also share chronos, but only one frequency would be recorded.

The EM clustering method described here relies on the assumption that clusters can be described by a Gaussian distribution. To check if the imposed Gaussian distributions significantly influence the cluster outcomes, we have also used the agglomerative hierarchical (AH) clustering algorithm \cite{Day:1984}  which does not make such an assumption. The initial condition for AH clustering is that each fluctuation structure defines a cluster. Using a suitable metric the two closest clusters are combined, iterating the process until we have $N_{Cl}=1$ gives a (prohibitively large) cluster tree. Compression of the AH cluster tree can be achieved by filtering out clusters with small populations, allowing for a visualisation similar to the EM cluster tree in figure \ref{fig:eg_cluster_tree}; a set of AH clusters which are essentially equivalent to the $N_{Cl} = 10$ level of the EM cluster tree are also shown in figure \ref{fig:eg_cluster_tree} . The clusters resulting from the EM and AH methods have been found to be essentially the same apart from the weakly defined clusters (EM:K,L) and the `remainder' (AH:Rem), a quantitative comparison between populations of EM and AH clusters in figure  	\ref{fig:eg_cluster_tree} is shown in table \ref{arr:EM_AH_matrix_tgt10ms}.

\begin{table}
\begin{center}
\begin{tabular}[c]{l|lllllllll||l}
Cluster & AH:A & AH:B & AH:C & AH:D & AH:E & AH:F & AH:G & AH:H & AH:(Rem.) & total\\
\hline
EM:H    & 307 &    &    &    &    &    &   &    & 1    & 308 \\
EM:I    & 152 &    &    &    &    &    &   &    &     & 152 \\
EM:E    &    & 161 &    &    &    &    &   &    & 1    & 162 \\
EM:F    &    & 3   & 155 &    &    &    &   &    & 10   & 168 \\
EM:O    &    &    &    & 88   &    &    &   &   & 40   & 128  \\
EM:M    &    &    &    &    & 50  &    &   & 28  & 78   & 156 \\
EM:J    & 3   &    &    &   &    & 40  &   &    & 52   & 95 \\
EM:N    &    &    &    &    &    &    & 36 &    & 21   & 57  \\
EM:K    & 33  & 3   & 6   & 16  &    &    &   & 5   & 289  & 352 \\
EM:L    & 25  & 2   & 3   &    &    &    &   &    & 392  & 422 \\
\hline
\hline
total & 520 & 169 & 164 & 104 & 50  & 40  & 36 & 33  & 884  & 2000
\end{tabular}
\caption{A comparison of populations of clusters produced by the EM ($N_{Cl} = 10$) and AH algorithms, clusters are shown in figure \ref{fig:eg_cluster_tree}}
\label{arr:EM_AH_matrix_tgt10ms}
\end{center}
\end{table}

It is important to consider the scalability and computational requirements of the algorithm. Given fixed values of $N_c$ and $\Delta t$, the size of $S$ remains constant and the preprocessing stage has complexity $\mathcal{O}(N_\mathcal{S})$, where $N_\mathcal{S}$ is the number of timeseries datasets $\mathcal{S}$. The scalability of the clustering stage depends on the algorithm used, for the EM case we have $\mathcal{O}(N_{Cl}N_\alpha)$, which gives $\mathcal{O}(N_\mathcal{S})$ for constant $N_{Cl}$. The AH clustering algorithm is less desirable as it has complexity $\mathcal{O}(N_\alpha^2)$ due to distance calculation between each pair of fluctuation structures.

We have implemented the preprocessing and visualisation stages using the python language with the Scipy and Matplotlib libraries \cite{pythonurls}. The preprocessing of our dataset, 4600 S arrays ($28$ by $1000$), takes around 2 hours using a $1.9\,$GHz Intel Pentium M processor. The results are stored in MySQL tables; a table of fluctuation structure properties excluding $\Delta\psi$-space mapping is around $5\,$Mb in size, with the $3.6\times10^6$ rows of the $\Delta\psi$ mapping table taking around $30\,$Mb, using optimal data types. For clustering, we have used the EM algorithm from the WEKA suite of data mining tools \cite{weka,wekaurl} which runs at about $0.05\times N_{Cl} \times N_\alpha$ CPU seconds using $2.2\,$GHz AMD Opteron processors. For each $N_{Cl}$, 100 randomised initial conditions were used; the results with maximal log-likelihood are selected as the best clusters. The WEKA algorithm does not operate with toroidal data, so we map the $\Delta\psi-$space from the $N_c$-dimensional torus to a $2N_c$-dimensional cube $[-1,1]^{2N_c}$ by taking the $\sin(\Delta\psi)$ and $\cos(\Delta\psi)$ components. For the present work, no specific efforts were made to optimize the clustering process; more efficient clustering routines exist, including, for example, genetic algorithms for faster convergence.

The physical nature of the fluctuations in our example dataset is not yet completely understood. The dependence of spectra on plasma density $n$ and $\iotabar$ suggests a dispersion relation similar to that of the global Alfv\'en eigenmode (GAE) \cite{Wong:1999,Spong:2003}. However, the observed frequencies are smaller than the expected GAE frequencies by a factor of around $1/3$ \cite{Pretty:2007}; an experimental campaign is presently being undertaken in order to resolve this difference.

\section{Conclusion}
We have presented a highly automated data mining process for the characterisation of fluctuations in multichannel timeseries data. The manual interaction is restricted to two tasks: the selection of a cross-power threshold $\gamma_{min}$ and the choice of appropriate filter parameters. The former requires initial consideration of threshold effectiveness on a small subset of data, while the latter is an operation applied to the dataset as a whole. 

Given an appropriate choice of clustering algorithm, the data mining process scales well, with complexity $\mathcal{O}(N_\mathcal{S})$. We have used the procedure here with magnetic fluctuation data from configuration scans in the H-1 heliac, identifying different modes in parameter space. The process should be easily adaptable to other types of multichannel oscillatory timeseries data.

\section*{Acknowledgements}
The authors would like to thank the H-1 team for continued support of experimental operations as well as J. Harris, F. Detering and M. Hegland for useful discussions. This work was performed on the H-1NF National Plasma Fusion Research Facility established by the Australian Government, and operated by the Australian National University, with support from the Australian Research Council Grant DP0344361 and DP0451960.

\end{document}